\documentclass{ifacconf}

\usepackage{graphicx}      
\usepackage{natbib}        
\usepackage{multirow}
\usepackage{array}
\usepackage{amsmath}
\usepackage{color}
\usepackage{amssymb,array,enumerate,epsfig,pifont,graphics,subfigure}
\newtheorem{definition}{Definition}
\newtheorem{rmk}{Remark}

\begin{document}

\begin{frontmatter}

\title{Perception and Steering Control in Paired Bat Flight\thanksref{footnoteinfo}} 

\thanks[footnoteinfo]{The authors gratefully acknowledge the funding support provided by ONR MURI grant N00014-10-1-0952 awarded to Boston University through the University of Washington.}

\author[*]{Zhaodan Kong} 
\author[*]{Kayhan \"Ozcimder} 
\author[**]{Nathan W. Fuller}
\author[***]{John Baillieul}

\address[*]{Department of Mechanical Engineering, Boston University \newline (email: \{zhaodan, kayhan\}@bu.edu)}                                              
\address[**]{Center for Ecology and Conservation Biology, Department of Biology, Boston University (email: nwfuller@bu.edu)}     
\address[***]{Department of Mechanical Engineering, Department of Electrical and Computer Engineering, Division of Systems Engineering,  Boston University (Corresponding author, email: johnb@bu.edu)}                               

\begin{abstract}                      
Animals within groups need to coordinate their reactions to perceived environmental features and to each other in order to safely move from one point to another. This paper extends our previously published work on the flight patterns of \emph{Myotis velifer} that have been observed in a habitat near Johnson City, Texas. Each evening, these bats emerge from a cave in sequences of small groups that typically contain no more than three or four individuals, and they thus provide ideal subjects for studying leader-follower behaviors. By analyzing the flight paths of a group of \emph{M. velifer}, the data show that the flight behavior of a follower bat is influenced by the flight behavior of a leader bat in a way that is not well explained  by existing pursuit laws, such as classical pursuit, constant bearing and motion camouflage. Thus we propose an alternative steering law based on \emph{virtual loom}, a concept we introduce to capture the geometrical configuration of the leader-follower pair. It is shown that this law may be integrated with our previously  proposed vision-enabled steering laws to synthesize trajectories, the statistics of which fit with those of the bats in our data set. The results suggest that bats use perceived information of both the environment and their neighbors for navigation.
\end{abstract}


\end{frontmatter}

\section{Introduction}

For a group of animals navigating through a cluttered environment, each individual must utilize sensory cues from both the environment and its neighbors in order to coordinate its motion with the neighbors and achieve effective navigation. A superb example of group navigation is bats emerging from their roost in large groups shortly after sunset and flying through a wooded flight corridor to their forage ground. \cite{kong2013optical} analyzed data recovered from a large collection of video records of a group of \emph{Myotis velifer} emerging from a cave on the Bamberger Ranch Preserve near Johnson City, Texas, focusing on their sensorimotor behavior with respect to environmental features. In this paper, we continue to analyze the same data set by considering the interactions between pairs of bats with the aim of establishing a unified view of bat navigation behavior.

Based on the species involved and the nature of the flight, paired-animal flight interactions have been mainly studied in the context of two situations: chasing and following. Chasing refers to the case in which a predator tries to catch a prey. \cite{mizutani2003insect} and \cite{ghose2006echolocating} show that bats and dragonflies use a motion camouflage flight strategy, which minimizes motion parallax cues that the prey can extract from its optical flow. Following is less aggressive than chasing and is generally conspecific. \cite{chiu2010effects} shows that a follower bat demonstrates such a behavior to conceal itself from the leader bat in order to increase its prey-capture performance. A classical pursuit strategy, in which the follower points its velocity vector towards the leader, is preferred in following. 

\begin{figure}[!tbh]
\centering
\includegraphics[width=\columnwidth]{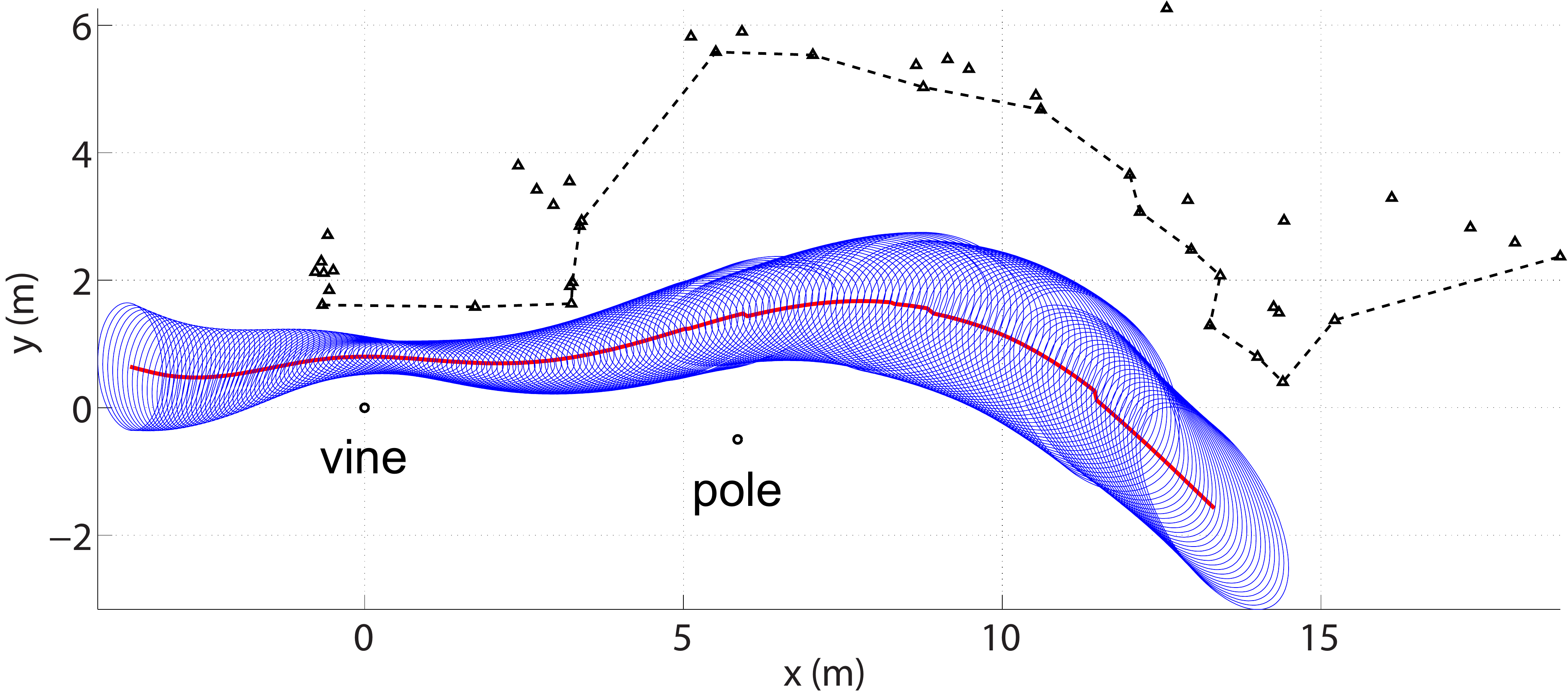}
\caption{Flight path statistics for 39 {\em M. velifer} are depicted. The red curve is the mean trajectory, and the blue ellipses (centered on the mean trajectory) represent a dispersion of one standard deviation. Two obstacles, a vine and a pole, are presented as circles. The triangles are visual features in a wooded area (mostly tree branches) and the dotted lines define the edges of the wooded area. See Appendix \ref{appen:computation} for a brief description on how the path statistics are computed.}
\label{fig:Variance1new}
\end{figure}

In \cite{sebesta2012animal} and \cite{kong2013optical}, we discussed the concept of \emph{time-to-transit} and used it as the basis for a collection of vision-based steering control laws. It was argued that time-to-transit was a biologically meaningful parameter that could probably be calculated in an animal's visual cortex, and steering control laws based on time-to-transit relative to single and pairs of environmental features were proposed. Our research assumed that the bats navigated through the flight corridor depicted in Fig. \ref{fig:Variance1new} by stitching together sequences of motion primitives in each of which visual feedback focused on either one or two environmental features. It was shown that even a very small set of such motion primitives was expressive enough to allow a simulated air vehicle to fly a bat-like trajectory. The keys to generating animal-like trajectories for a simulated flight vehicle were: 
\begin{itemize}
\item designing a set of vision-based motion primitives that produce motion segments based on the geometry and movement of
image points on the focal plane (retina) of the image sensor;
\item sequentially updating a set of key feature points and selecting the appropriate motion primitives to guide the vehicle along each path segment;
\item a protocol for switching between the key features of one segment and the next. 
\end{itemize}

It was observed that along those portions of the flight corridor where environmental clutter was relatively dense, each of the motion segments needed to be focused on closely spaced features and was of short duration. Along these portions of the motion, switching between control laws (and features) was frequent (e.g. near the vine in Fig. \ref{fig:Variance1new}). Along portions of the flight path where there was less clutter, the animal-like motion segments appeared to use feedback based on more widely separated features. There was thus a moderating effect such that the simulated flight path took only a shallow excursion toward the concave edge of the wooded area, rather than following the edge more closely and at a constant distance as would have been the case if the steering laws were based on rapid updates of closely spaced features. Comparing the simulated flight paths based on such considerations with the paths reconstructed from animal field data, we developed the hypothesis that the animal movements were guided by both direct reaction to environmental features and some form of cognitive processing that could involve spatial memory and path choices to minimize energy expenditures. Using our control primitives, we were able to develop a motion strategy that would closely approximate the mean flight path of the bats (the red curve in Fig. \ref{fig:Variance1new}). The question remained as to why many animals deviated significantly from this mean path. In the present paper, we propose that large excursions toward the boundary of the woods could be the result of a trailing bat following a leader according to a certain leader-follower protocol. Using the concept of \emph{virtual loom}, we formulate a new steering law that produces simulated flight paths consistent with those of pairs of bats observed in the field.

The paper is organized as follows. Section \ref{section:tau} introduces the concept of virtual loom. Section \ref{section:statistical} presents our analysis of \emph{M. velifer}'s apparent following behavior and tests the data against existing pursuit laws. Section \ref{section:steering_law} proposes a {\em virtual loom} based steering law that produces the kinds of following behaviors that have been observed in bat pairs in the field. In Section \ref{s:simulation}, we describe motions of simulated vehicles that use synthetic images of both stationary features and moving objects (a leader bat or another vehicle) to guide motion through the computer reconstructed flight corridor. A number of vehicle simulations were carried out with vehicles entering the flight corridor at random (Poisson) times and random (Gaussian) locations across the left hand boundary of the flight corridor. Comparisons with our bat flight data are made and show that the simulations have significant similarities. Next steps in the research are discussed in Section \ref{s:sconclusion}.

\section{Virtual Loom}
\label{section:tau}

We model flight kinematics following the model of \cite{justh2006steering}. The dynamics of the leader are given as:
\begin{equation}
\label{e:leader}
\left \{
\begin{array}{ll}
\dot{\boldsymbol{r}}_l &= v_l \boldsymbol{x}_l\\
\dot{\boldsymbol{x}}_l &= v_l \boldsymbol{y}_l u_l\\
\dot{\boldsymbol{y}}_l &= -v_l \boldsymbol{x}_l u_l,
\end{array} \right.
\end{equation}
where $v_l$ is the speed of the leader, $\boldsymbol{r}_l$ is the position of the leader, $\boldsymbol{x}_l$ is the unit tangent vector to the trajectory of the leader, $\boldsymbol{y}_l$ is the corresponding unit normal vector, and the plane curvature $u_l$ is the steering control for the leader. Similarly, the dynamics of the follower are given as:
\begin{equation}
\label{e:follower}
\left \{
\begin{array}{ll}
\dot{\boldsymbol{r}}_f &= v_f \boldsymbol{x}_f\\
\dot{\boldsymbol{x}}_f &= v_f \boldsymbol{y}_f u_f\\
\dot{\boldsymbol{y}}_f &= -v_f \boldsymbol{x}_f u_f.
\end{array} \right.
\end{equation}
In this paper, we assume that the leader and the follower have the same speed.

\begin{figure}[!tbh]
\centering
\includegraphics[width=0.8\columnwidth]{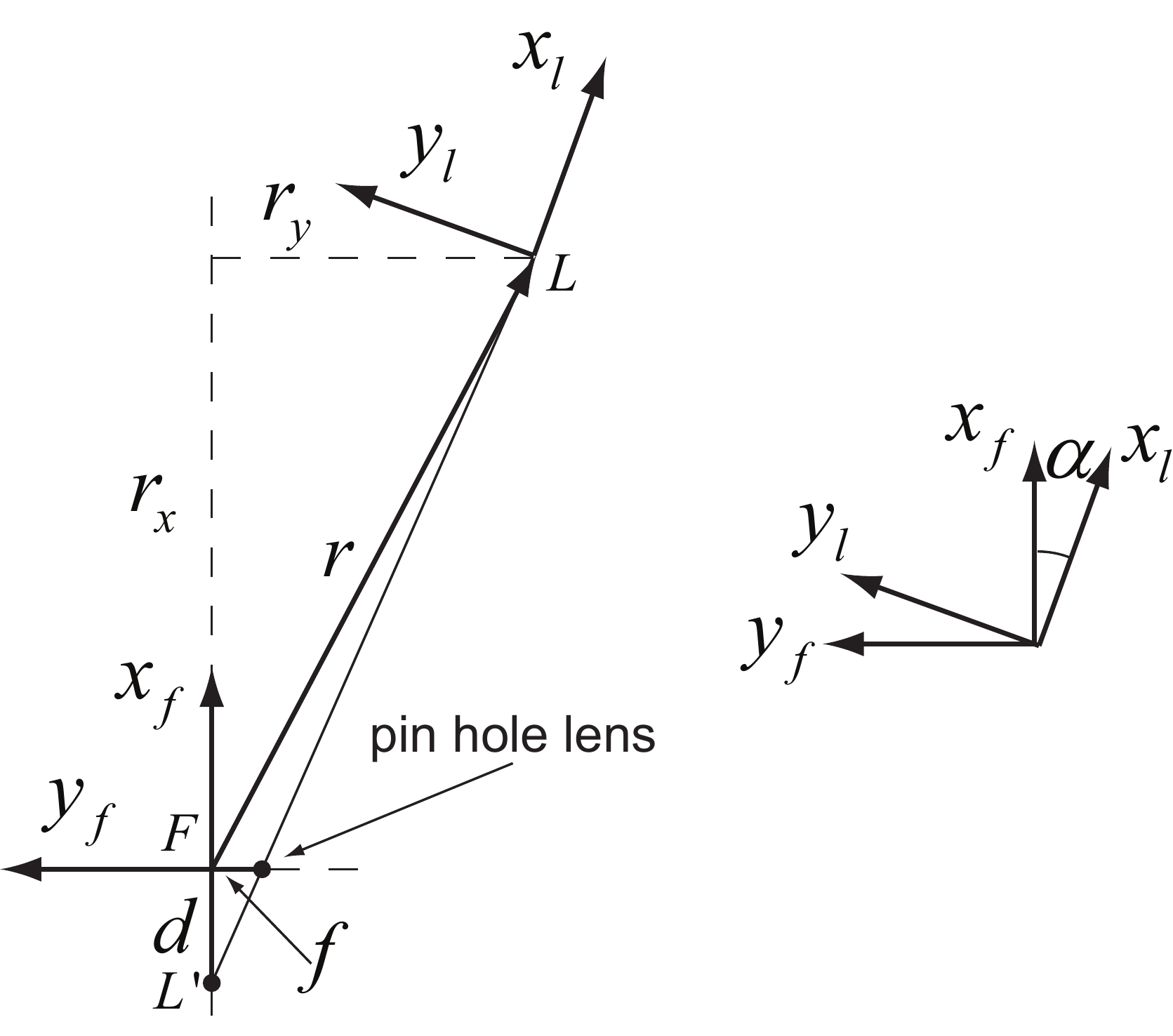}
\caption{Frenet frame representations of the leader and the follower together with the follower's side-looking system. $L$ and $F$ are the center axis points of the eyes of the leader and the follower, respectively. $|f|$ is the focal length distance from the lens to the focal plane (retina). $L^{\prime}$ is the image point corresponding to $L$. $\alpha$ is the angle between $\boldsymbol{x}_l$ and $\boldsymbol{x}_f$.}
\label{fig:Planar_Motion}
\end{figure}

Fig. \ref{fig:Planar_Motion} illustrates the geometry related to an idealized leader-follower pair moving in a horizontal plane\footnote{As noted in \cite{kong2013optical}, the bat motions in our data set are approximately planar.}. The directions of motion are aligned with the vehicle body frame x-axes, i.e., $\boldsymbol{x}_l$ and $\boldsymbol{x}_f$. The leader is observed by the follower with a pinhole camera system whose camera axis is aligned with the follower's negative body frame y-axis, i.e., $\boldsymbol{y}_f$. The relative position of the leader in the frame of the follower is $\boldsymbol{r} = \boldsymbol{r}_l-\boldsymbol{r}_f$. The projections of $\boldsymbol{r}$ onto the $\boldsymbol{x}_f$ and  $\boldsymbol{y}_f$ directions are written as:
\begin{equation}
\label{e:rx_ry}
r_x := \boldsymbol{r} \cdot \boldsymbol{x}_f \text{ and } r_y := \boldsymbol{r} \cdot \boldsymbol{y}_f
\end{equation}
respectively. 

In terms of these kinematic models and the follower's imaging system model that is depicted in Fig. \ref{fig:Planar_Motion}, we recall definitions of optical flow parameters from \cite{sebesta2012animal} and \cite{kong2013optical}. If the follower position at some initial time $t=0$ is $\boldsymbol{r} _f(0)=(r_1(0),r_2(0))$ with the leader being stationary and the follower flying in its body frame direction $\boldsymbol{x}_f$ at a constant speed $v_f$, it will cross the {\em line of transit} that is perpendicular to the line of flight and passes through the origin of the leader frame at time $\tau=\boldsymbol{r}_x/v_f$, where $\boldsymbol{r}_x$ is the distance between $\boldsymbol{r}_f(0)$ and this same line of transit. This quantity is called the {\em time-to-transit}, and we denote it by $\tau$. It has been widely studied in literature dealing with motion perception (See e.g. references in \cite{sebesta2012animal}.), and it has been shown to be easily computed in an animal's visual cortex. Indeed, if, at the initial time ($t=0$), $d$ is the distance in the follower's image plane (bat retina) between the leader's image ($L^{\prime}$ in Fig. \ref{fig:Planar_Motion}) and the principal camera axis point $F$ (Fig. \ref{fig:Planar_Motion}), then $\tau=d/\dot d$. If the leader is not stationary, the definition still makes sense and is related to the relative velocities of the leader-follower pair. Of course if the leader and follower are traveling in the same direction at the same speed, the image distance $d$ does not change over time ($\dot d=0$), which reflects the fact that $\tau$ must be infinite. Since we shall be interested largely in the case where our leader and follower fly at essentially identical speeds, we find it more convenient to work with the reciprocal of $\tau$, which is called the {\em loom}. Since we shall be dealing in particular with situations in which the follower never reaches the point of transit, we define the \emph{virtual loom} as follows:

\begin{definition}
For a leader-follower pair (Eqs. (\ref{e:leader}) and (\ref{e:follower})), the \emph{virtual loom} $\lambda(t)$ at time $t$ is
\begin{equation}
\label{e:def_tau}
\lambda(t) = \frac{[1-\boldsymbol{x}_f (t) \cdot \boldsymbol{x}_l (t)] v_f}{\boldsymbol{r}(t) \cdot \boldsymbol{x}_f(t)}.
\end{equation} 
Notice that $\lambda(t)$ has a unit that is inverse of time. For brevity, we use $\lambda$ to represent $\lambda(t)$.
\end{definition}

From Fig. \ref{fig:Planar_Motion}, we have the following relationship:
\begin{equation}
d = \frac{f}{r_y-f} r_x,
\end{equation}
so the follower bat can estimate $r_x$ by sensing $d$.

In addition, we define an equilibrium state for a pair as follows.
\begin{definition}
A leader-follower pair (Eqs. (\ref{e:leader}) and (\ref{e:follower})) is said to be in a state of \emph{$\lambda$ equilibrium} if $\lambda$ is zero.
\end{definition}

\begin{rmk} 
\label{r:bio}
Suppose, as shown in Fig. \ref{fig:Planar_Motion}, $\alpha$ is the angle between the headings of the two bats, then $\cos \alpha = \boldsymbol{x}_f \cdot \boldsymbol{x}_l$. Further, define \emph{transiting} as the instant when the image of the leader on the follower's retina $L'$ coincides with $F$, the focal point of the follower's retina, which corresponds to $r_x=\boldsymbol{r} \cdot \boldsymbol{x}_f=0 $. For two bats flying with the same constant speed $v_f=v_l=v$, a state of $\lambda$ equilibrium means that the relative velocity of the two bats is zero and $L'$ stays at the same position on the follower's retina. In this case, the follower bat can estimate $\alpha$ by sensing $\dot{d}$, the optical flow. A zero $\dot{d}$ corresponds to a zero $\alpha$. On the other hand, a non-zero $\dot{d}$ implies that $\alpha$ is not zero and a transiting is going to happen if no adjustment is made by the follower. Finally, it is worth pointing out that, although in this paper we focus on vision-based control, bats can also use other sensory modalities, such as echolocation (\cite{shaw1991acoustic}), to estimate time-to-transit $\tau$ or virtual loom $\lambda$. 
\end{rmk}

\begin{rmk} 
\label{r:parallel}
Parallel (or near parallel) flight alignment (with $\alpha \cong 0$) has been observed in the mating activity of dragonflies (\cite{wagner1986flight}), competitive prey capturing in bats (\cite{chiu2010effects}) and tandem flight of swallows (unpublished results from The Hedrick Lab at UNC Chapel Hill). Benefits of such a flight pattern include aerodynamic efficiency (the follower can utilize the vortex of the leader's wingtip to save energy, known as `vortex surfing' (\cite{lissaman1970formation})), stealth (the follower can conceal itself from the leader to increase its prey capturing probability) and echolocation efficiency (the follower bat can turn off its sonar or adopt a low duty cycle). 
\end{rmk}

\section{Flight Behavior of Myotis Velifer: Data Analysis}
\label{section:statistical}

In this section, we describe the experiment procedure regarding the data collection on flight behavior of \emph{M. velifer} and then the analysis results. 

\subsection{Experiment Procedure}

Raw bat flight data were collected shortly after sunset on 30 May, 2011. The bat colony resides in an artificial cave located approximately 50 meters from the point of observation. Upon exiting the roost, individuals immediately begin to disperse over the landscape by following the margin of a forest fragment toward an open flight corridor over a paved ranch road. We collected thermal infrared video of bats with three thermal cameras (FLIR ThermoVision SC8000, FLIR Systems, Inc.) placed along the flight corridor. We chose this location because there was an abundance of natural obstacles in the flight corridor and because it was sufficiently far from the roost that the bats presumably had accreted into flight groups but were not sufficiently far from the roost to have split from each other to forage separately. Our camera system operated at 131.5 Hz with 1024$\times$1024 resolution and used 25 mm lenses. 

Cameras were placed linearly and perpendicular to the flight direction of the bats. Camera viewing angles were selected so as to optimize reconstruction accuracy at points of direct interaction between bats and a natural obstacle (a hanging vine), and to maximize flight track duration. On average, each bat was recorded for approximate 300 frames. This was accomplished by localizing the vine at a central focal point in each of the three camera views. The 3D geometry of the scene was calibrated by waving an object of known dimension through the shared view volume of the three cameras, in this case a 1.56 m PVC ``wand", and direct linear transformation (DLT) coefficients were calculated from pairs of wand points (for more information refer to \cite{abdel1971direct}). A technician gathered 2D coordinates of each bat in each of the three views using custom annotation software developed by our research group. Flight trajectories were then reconstructed in 3D as described in \cite{towne4error}. For hand-annotated positions, some human-generated noise was introduced to the flight trajectories. This uncertainty was smoothed as described in Appendix \ref{appen:computation}.

\subsection{Poisson Emergence}

\begin{figure}[!htb]
\centering
\includegraphics[width=.9\columnwidth]{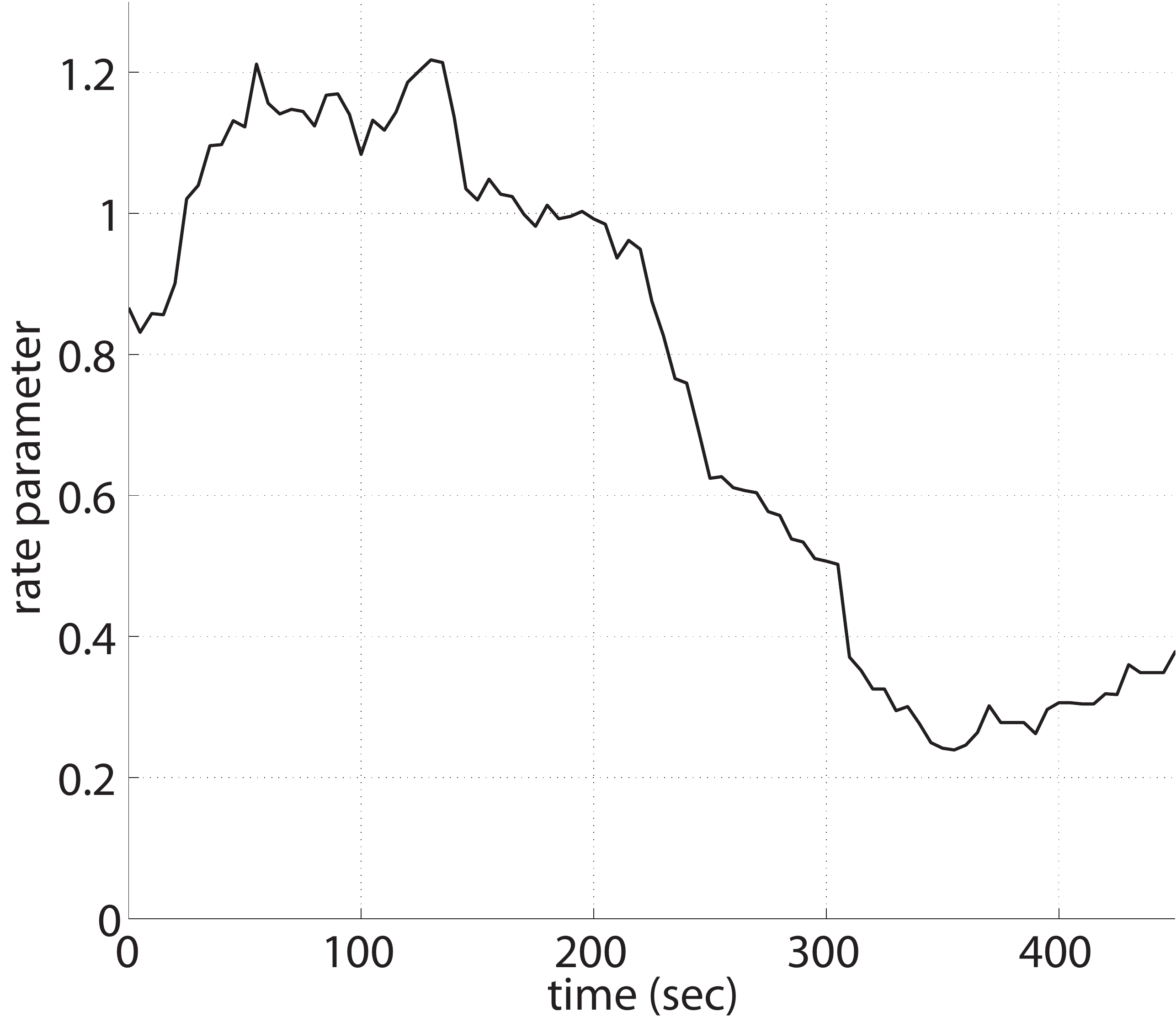}
\caption{The sampled rate parameter $\bar{\theta}(t)$ of the subset $\{t_i \in [t,t+T]\}$ with $t \in [0, 450]$ and $T=120$ seconds. The time axis corresponds to the whole duration of the recording period with 0 corresponding to the time the recording started.} \label{fig:Poisson}
\end{figure}

Previous study of bat emergence time has been largely focused on how factors, such as sunset time, weather and existence of predators, affect the onset of the emergence and the mean emergence time (\cite{welbergen2006timing,kunz1996variation}). To our knowledge, there has been no study to model the fine details of emergence rates. However, there exists a rich set of literature on the modeling of human activity emergence, such as sending emails and initiating financial transactions (\cite{barabasi2005origin}). 

We define the first time a bat appears in the video as its emergence time. By this, we get an ordered time sequence, $S:=\{t_i,i=1,...,N\}$, where $t_i$ is the emergence time of the $i$th bat and $N=254$ is the total number of recorded bats. Notice that Fig. \ref{fig:Variance1new} only shows a fraction of the trajectories. Please refer to \cite{kong2013optical} for details regarding the whole data set. The Kolmogorov-Smirnov test (\cite{lilliefors1969kolmogorov}) is used to determine whether the sequence (or a subset of it) fits a Poisson model. 

Fig. \ref{fig:Poisson} shows the sampled rate parameter $\bar{\theta}(t)$ of the subset of emergence times that fall within the window $[t,t+T]$. It can be seen that $\bar{\theta}(t)$ is relatively constant before 200th second and its value is high; it falls rather sharply after 200th second; it becomes relatively constant again after 300th second. Our analysis has shown that the entire emergence time sequence $S$ does not fit a Poisson model. However, the analysis also has shown that the truncated emergence time sequence $S_1:=\{t_i \in [0, 200]\}$ is able to pass the Kolmogorov-Smirnov test for a Poisson arrival process with a constant rate parameter $\bar{\theta}$ of 0.961\footnote{Another truncated sequence $S_2:=\{t_i \in [300, 450]\}$ was also tested. But it did not pass the test due to the the lack of enough data points for statistical significance.}. Further, it has been found that bats emerging within the duration $[0, 200]$ account for 80 percent of the bats.

If we look at an interval of one second, a Poisson arrival process with a rate parameter 0.961 means that there is a 0.3825 probability that there is no bat within the interval, a 0.3676 probability that there is one bat within the interval, a 0.2499 probability (approximately 64 bats for sample of 254) that there are two or more bats within the interval. Due to the high probability of having neighboring bats, in the next subsection, we will study whether the behavior of a leader bat affects the behavior of a follower bat and if it does, in what way.

\subsection{Effects of Leader on Follower}
\label{sec:effects}

\begin{figure}[!tbh]
\centering
\includegraphics[width=\columnwidth]{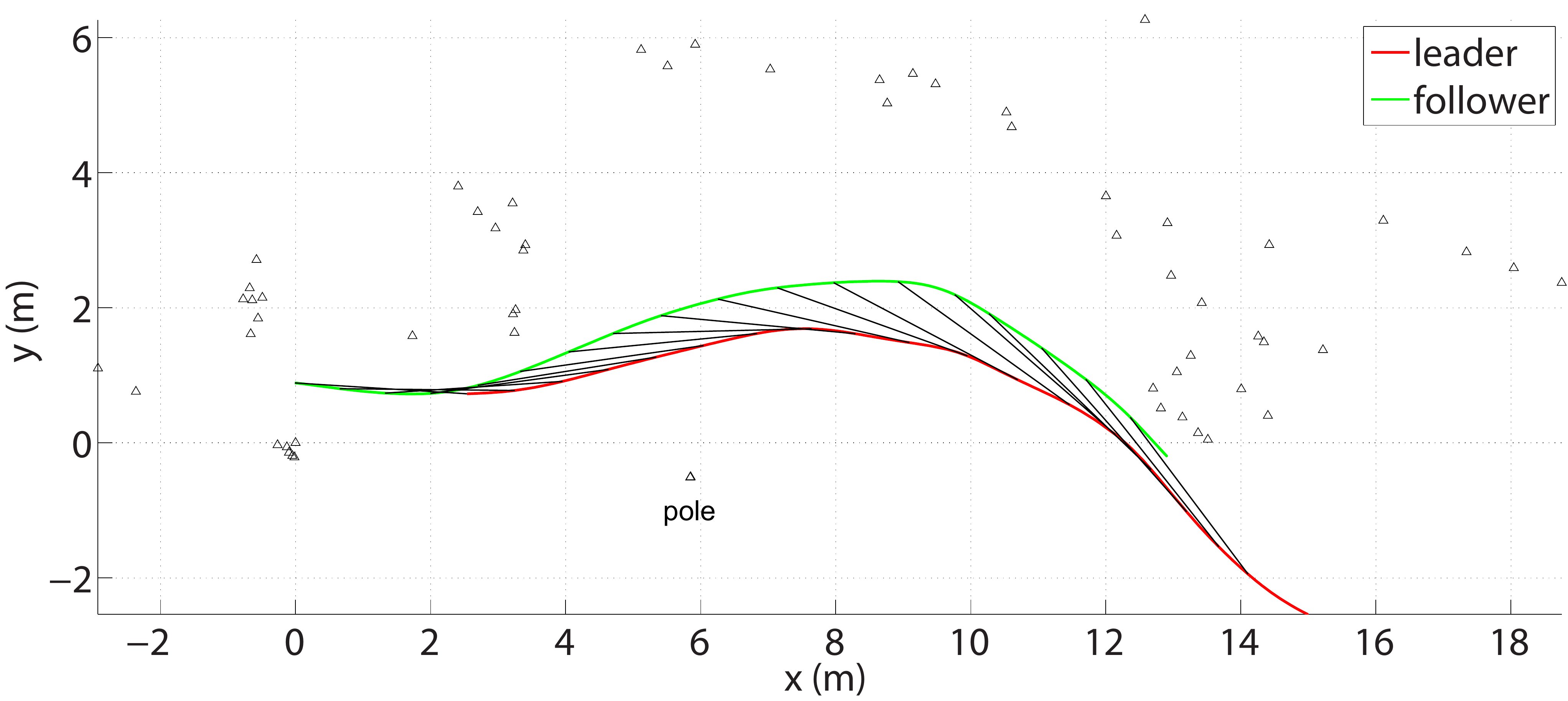}
\caption{Example trajectories of a leader-follower pair. The black lines connect the paired bats' corresonding locations at different time slices.} \label{fig:Original_Trajectory}
\end{figure}

\begin{figure}[!tbh]
\centering
\includegraphics[width=.9\columnwidth]{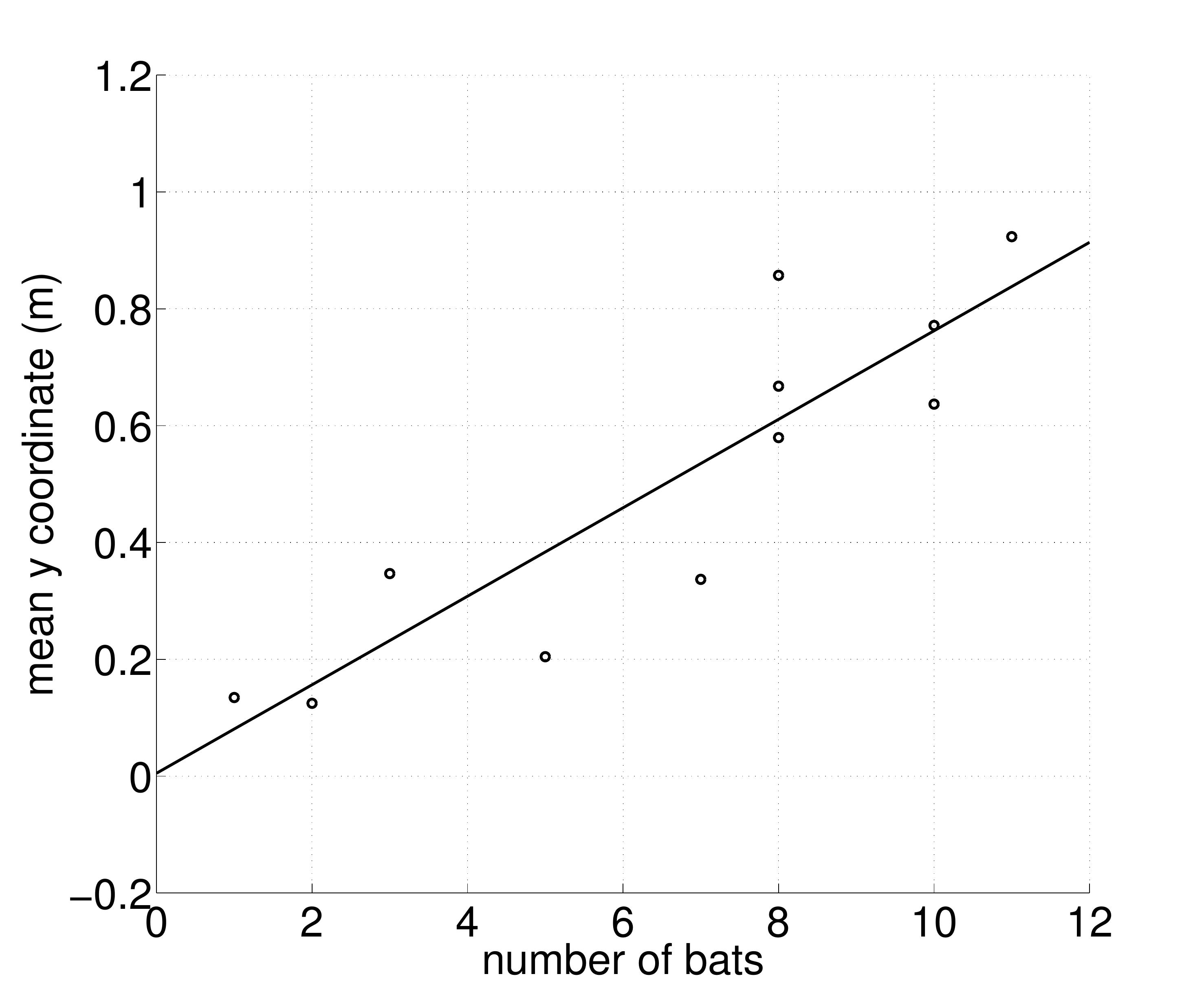}
\caption{As explained in the text, there is a relationship between the mean $y$ coordinates of bats emerging within a 40 second time window and the number of bats in the window ($R=0.8894$).} \label{fig:Relationship_Number}
\end{figure}

We classify the 254 collected trajectories into six groups based on their positions with respect to the obstacles (the vine and the pole). The 39 trajectories that are shown in Fig. \ref{fig:Variance1new} and will be analyzed in this subsection correspond to the group of bats passing the vine from the left and passing the pole from the left while flying lower than the upper end of the pole. There are other groups, such as those passing the vine from the right and passing the pole from the right. See \cite{kong2013optical} for the information regarding the classification and other groups.

For the group of 39 trajectories shown in Fig. \ref{fig:Variance1new}, we further select data segments for analysis based on the following criteria: the paired bats need to appear in the video simultaneously for longer than 20 frames and the spatial separation between the paired bats must be shorter than 10 meters\footnote{A bat can perceive items within 10 meters with a good resolution via its eyes (\cite{wimsatt1970biology}). Given that the average speed of the observed bats is 10.17 m/s, this threshold corresponds to approximately one second difference between the two bats' emergence times.} . We say that the bat emerging earlier is the leader and the one emerging later is the follower. The trajectories of one such pair are shown in Fig. \ref{fig:Original_Trajectory}.

From our analysis of the data, we find a list of statistically significant relations (measured by the Pearson correlation coefficient $R$):
\begin{itemize}
\item There is a correlation of $R=0.8894$ between the mean $y$ coordinate of bats emerging within a fixed time window and the number of bats in the window. The result is shown in Fig. \ref{fig:Relationship_Number}. As shown in Fig. \ref{fig:Original_Trajectory}, a higher $y$ coordinate implies a closer distance to the woods. Further, the larger the number of bats emerging within a fixed time window, the shorter the average interval between the successive emergence of two bats, and the higher the probability of having a leader in front of a bat. 
\item For pairs of bats, there is a correlation of $R=-0.5104$ between the difference of the mean $y$ coordinates of the follower and the leader and their initial distance from each other. Combined with the above relation, this relation implies that a follower bat tends to fly closer to the wooded area than its leader.
\item For pairs of bats, there is a correlation of $R=-0.4589$ between the difference of the mean route lengths of the follower and the leader (with the two routes covering the same $x$ range [0, 12]) and their initial distance form each other. The relation implies that a follower bat tends to take a longer route than its leader.
\end{itemize}

\begin{figure}[!tbh]
\centering
\includegraphics[width=\columnwidth]{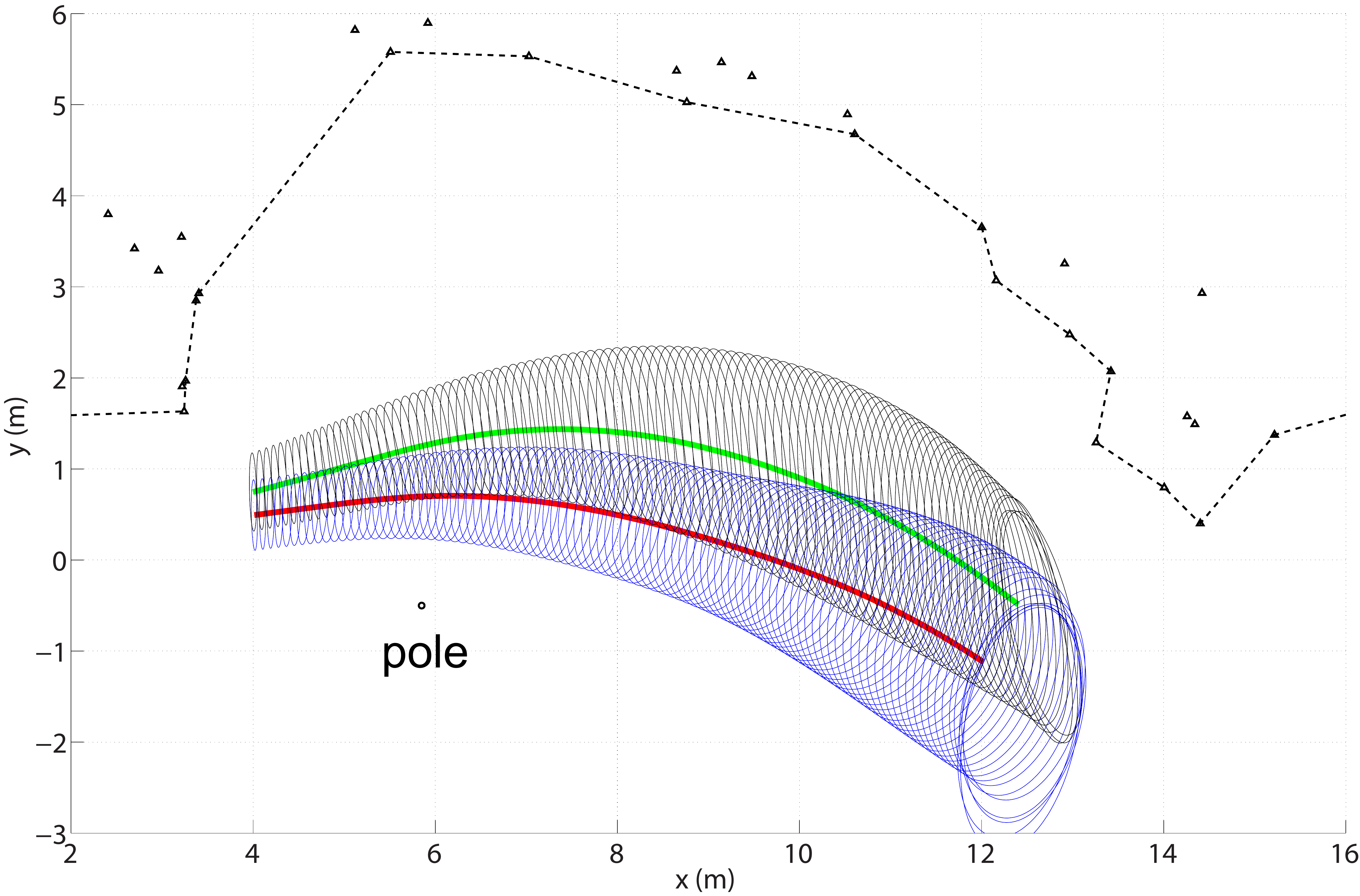}
\caption{Flight path statistics of $G_1$ and $G_2$ are depicted. The red (green) curve is the mean trajectory of $G_1$ ($G_2$). The blue (black) ellipses (centered on the mean trajectory) represent a dispesion of one standard deviation of $G_1$ ($G_2$).} \label{fig:leaderfollowerellipse}
\end{figure}

\begin{table}[!tbh]
\label{table:classification}
\caption{{\small Number of Bats in Each Class}}
\begin{center}
    \begin{tabular}{| c | c | c | c |}
    \hline\hline
     $C_1$ & $C_2$ & $C_3$ & $C_4$ \\ \hline
     7 & 14 & 4 & 14\\ \hline\hline
    \end{tabular}
\end{center}
\end{table}

These correlations mean that a bat (a follower) behaves differently if there is another bat (a leader) in front of it. In order to further illustrate the behavior difference, we classify the 39 trajectories shown in Fig. \ref{fig:Variance1new} into four classes. They are
\begin{itemize}
\item $C_1$: the bat is a single bat, which is neither a leader nor a follower;
\item $C_2$: the bat is a single-role leader bat, which is a leader but not a follower;
\item $C_3$: the bat is a dual-role bat, which is both a leader and a follower;
\item $C_4$: the bat is a single-role follower bat, which is follower but not a leader.
\end{itemize}
The numbers of bats in different classes are shown in Table \ref{table:classification}. We then combine the four classes into two groups: the leader group $G_1=\{C_1, C_2\}$ and the follower group $G_2=\{ C_3, C_4\}$. The statistics of the two groups are shown in Fig. \ref{fig:leaderfollowerellipse}. It is quite obvious that the follower group curves more toward the wooded area than the leader group.

To conclude, as the number of bats emerging within an interval becomes larger or equivalently the initial distance (the emergence interval) between the leader-follower pair becomes smaller, the follower bat tends to stay closer to the wooded area and take a longer route than the leader bat. One possible interpretation of the observed effects is that the trailing bat tries to maintain a relatively constant distance from the leader while staying a safe distance away from the obstacles, e.g. the pole. For the specific environment as shown in Fig. \ref{fig:leaderfollowerellipse}, a side effect of such a behavior is a larger excursion towards the woods for the follower bat. 

\begin{rmk} 
\label{r:new_data}
The analysis in this subsection is based on 39 bat trajectories, which were collected in a single day for around 8 minutes. Recently we have collected a much larger data set of the same species at the same location. The recording was 45 minute long for each day and lasted 8 days. We plan to perform the same analysis on the new data set to test and validate these observations.
\end{rmk}

\subsection{Are Bats Pursuing One Another?}
\label{sub:are}

\begin{figure}[!tbh]
\centering
\includegraphics[width=.9\columnwidth]{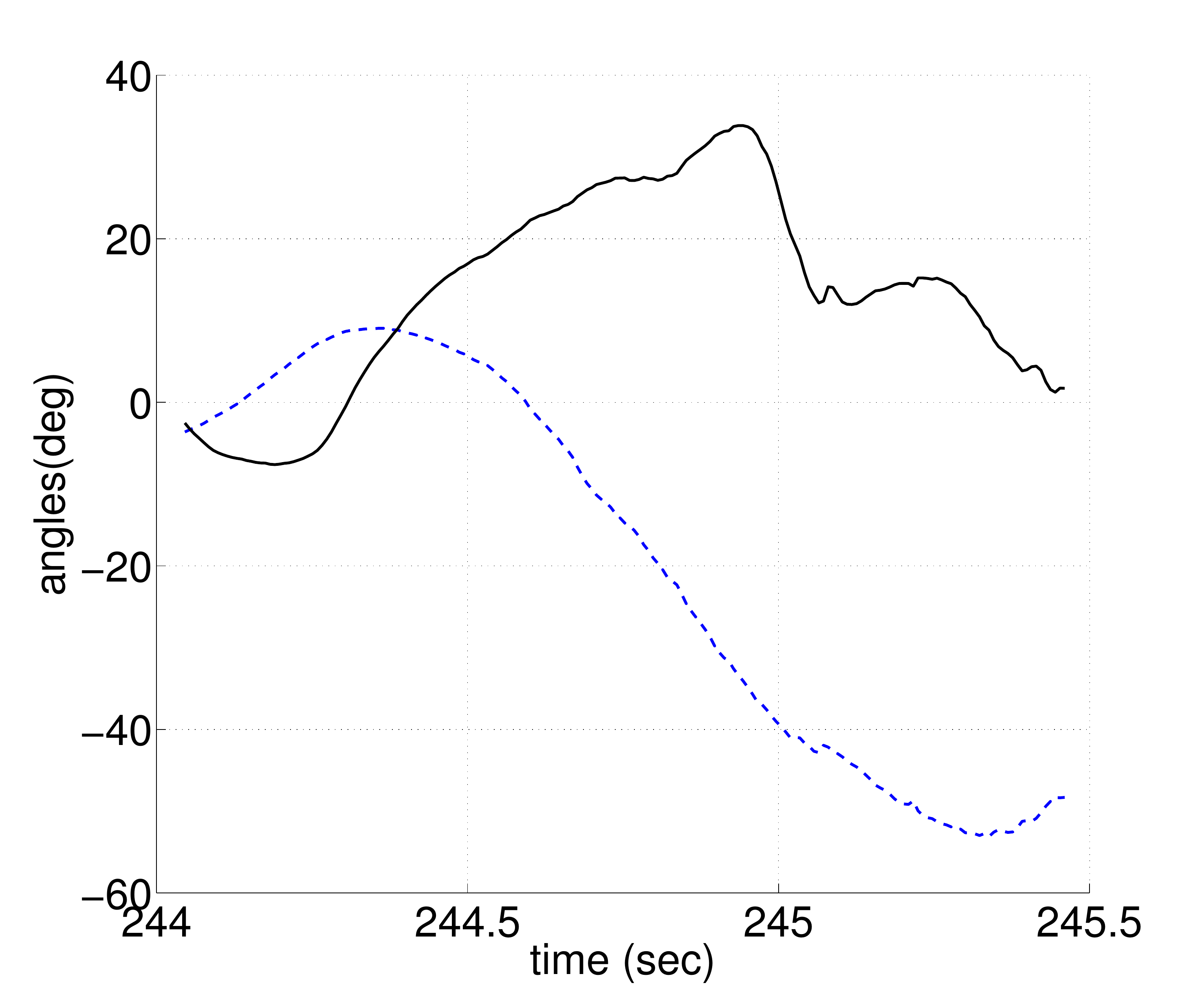}
\caption{Analysis results for the pair shown in Fig. \ref{fig:Original_Trajectory}: the baseline direction $\boldsymbol{r}/|\boldsymbol{r}|$ (blue) and the angle between the baseline direction $\boldsymbol{r}/|\boldsymbol{r}|$ and the follower's heading $\boldsymbol{x}_f$ (black). Both are represented as angles. For instance, the blue curve is computed by $\tan^{-1}(\boldsymbol{p}_{2}/\boldsymbol{p}_{1})$ with $\boldsymbol{p}_{1}$ and $\boldsymbol{p}_{2}$ as the first and second component of $\boldsymbol{r}/|\boldsymbol{r}|$.} \label{sub:Angle_Sequence}
\end{figure}

In this subsection, we analyze paired bats' behavior by checking the data against existing pursuit laws: classical pursuit, constant bearing and motion camouflage (\cite{wei2009pursuit}). In classical pursuit, the follower aligns its direction of motion $\boldsymbol{x}_f$ with the baseline direction $\boldsymbol{r}/|\boldsymbol{r}|$, where baseline $\boldsymbol{r}$ is defined as $\boldsymbol{r}_l-\boldsymbol{r}_f$ in Section \ref{section:tau}; in constant bearing, the follower keeps the angle between its heading $\boldsymbol{x}_f$ and the baseline direction $\boldsymbol{r}/|\boldsymbol{r}|$ constant; in motion camouflage, the follower keeps the baseline direction $\boldsymbol{r}/|\boldsymbol{r}|$ constant. 

Fig. \ref{sub:Angle_Sequence} illustrates that none of these pursuit laws explains the behavior observed in Fig. \ref{fig:Original_Trajectory}. The baseline direction $\boldsymbol{r}/|\boldsymbol{r}|$ (blue curve) does not stay constant, which violates motion camouflage pursuit; the angle between the baseline direction $\boldsymbol{r}/|\boldsymbol{r}|$ and the follower's heading $\boldsymbol{x}_f$ (black curve) is neither zero nor constant, which violates classical and constant bearing pursuits. The result implies that the follower bats are not pursuing the leader bats (by pursuing we mean that there is a moment when the follower intercepts the leader and they exchange their roles). The reasons may be as stated in Remark \ref{r:parallel}. Nevertheless an alternative interpretation is needed to explain the observed behavior. In the next two sections, we will propose a steering law and a navigation strategy the follower bat might use.

\section{Steering Law For Following}
\label{section:steering_law}

In this section, we propose a steering law that a trailing bat might use to follow another bat. Recall that we assume that the leader and the follower have the same speed.

\subsection{$\lambda$-Based Steering Law}

The planar steering law we study next is based on minimizing the virtual loom in a follower's perception of the leader's motion.
\begin{thm} 
\label{theorem2}
Consider leader-follower pair (Eqs. (\ref{e:leader}) and (\ref{e:follower})) with the following assumptions:
\begin{enumerate}
\item the control of the leader $u_l$ is zero (the leader flies in
a straight line);
\item $r_x$ is positive (the leader is in front of the follower).
\end{enumerate}
Then for $k>0$, the follower with control
\begin{equation}
\label{e:thm_control_tau}
u_f=k \boldsymbol{x}_l \cdot \boldsymbol{y}_f = -k \sin \alpha
\end{equation}
will asymptotically align itself with the leader, i.e., \ $\alpha \rightarrow 0$ (and $\lambda \rightarrow 0$).
\end{thm}

\begin{pf} 
We take the unnormalized virtual loom as a Lyapunov
function $V:= 1-\boldsymbol{x}_l\cdot\boldsymbol{x}_f$. This is $0$ if $\boldsymbol{x}_l
\cdot \boldsymbol{x}_f=1$ ($\alpha =0$) and positive otherwise. Its derivative along trajectories is
\begin{equation}
\label{proof_tau_2}
\begin{array}{rl}
\dot{V} &= -\dot{\boldsymbol{x}}_l \cdot\boldsymbol{x}_f-\boldsymbol{x}_l \cdot\dot{\boldsymbol{x}}_f\\
& = -u_f(\boldsymbol{x}_l\cdot\boldsymbol{y}_f)\\
& = - k(\boldsymbol{x}_l\cdot\boldsymbol{y}_f)^2
\end{array}
\end{equation}
which is zero when $\boldsymbol{x}_l \cdot \boldsymbol{y}_f = 0$ or equivalently $\boldsymbol{x}_l \cdot \boldsymbol{x}_f = 1$ ($\alpha = 0$).
\end{pf}

Theorem \ref{theorem2} implies that if the leader is flying in a straight line, then the follower can utilize the virtual loom to achieve parallel flight with the leader. See Remark \ref{r:bio} for the explanation of how bats might estimate the virtual loom. 

\subsection{Simulation Result}

\begin{figure}[!tbh]
\centering
\includegraphics[width=.9\columnwidth]{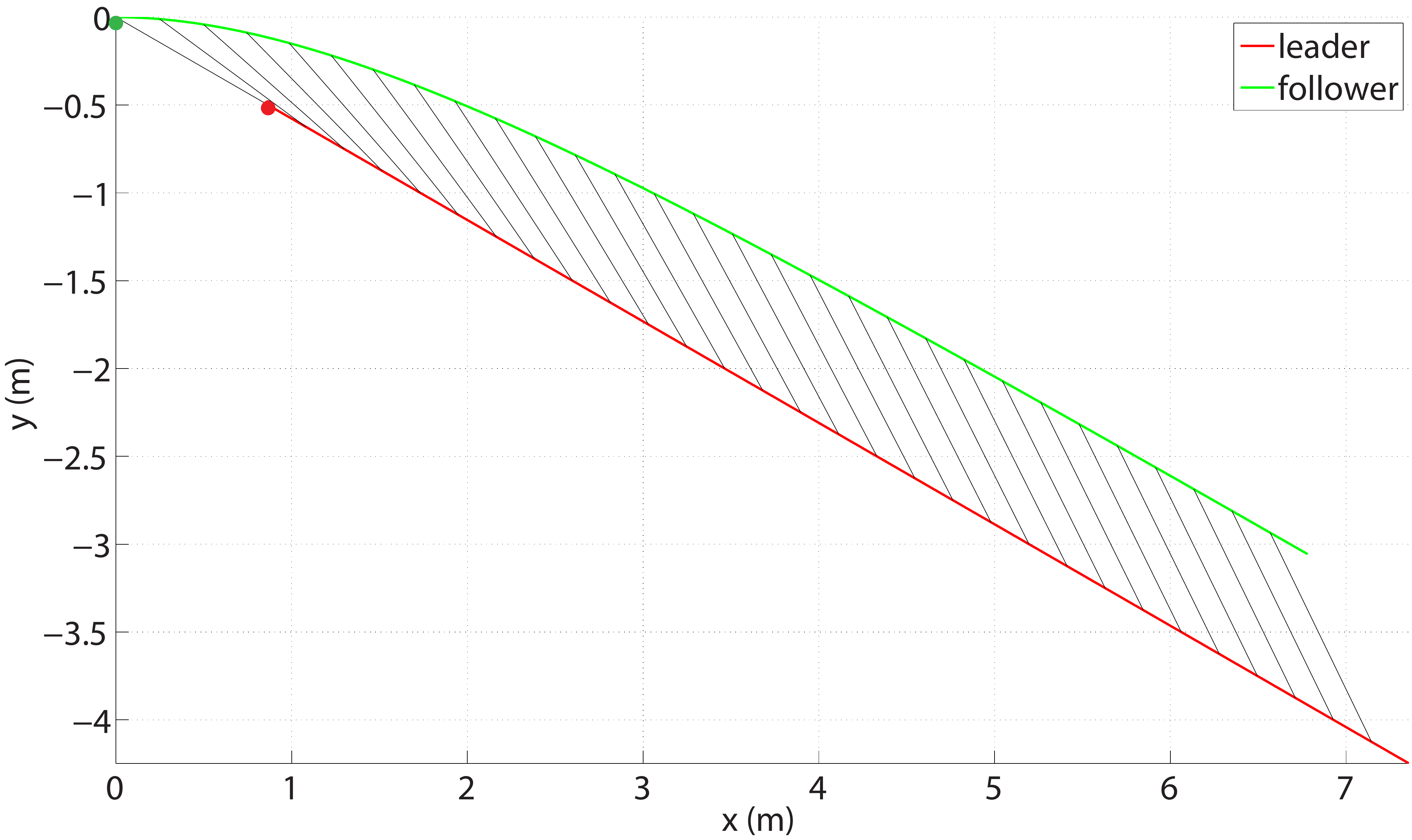}
\caption{Synthesized trajectories with the follower using control law (\ref{e:thm_control_tau}). Dots indicate starting locations. The black lines connect the pair's corresponding locations at different time slices.} \label{sub:Straight_1}
\end{figure}

Fig. \ref{sub:Straight_1} shows a pair of synthesized trajectories with the follower using control law (\ref{e:thm_control_tau}) and leader with the steering control $u_l = 0$. It can be seen that with control law (\ref{e:thm_control_tau}), the follower is approaching a parallel flight with the leader as described by Theorem \ref{theorem2}. In the case that $u_l \neq 0$, the leader's trajectory is similar to the one depicted in Fig. \ref{fig:Original_Trajectory}, and the synthesized follower's trajectory with control law (\ref{e:thm_control_tau}) is qualitatively similar to the actual follower bat's trajectory as shown in Fig. \ref{fig:Simulation_Data}. (Details will be elaborated in Section \ref{s:pure_following})

\section{Integrated Navigation Strategy}
\label{s:simulation}

In \cite{kong2013optical}, we proposed an integrated strategy to explain navigation behavior of \emph{M. velifer} in a data set of 254 individuals. We hypothesized that these bats used landmarks recalled from their spatial memory to select features from the environment and then generated control strategies based on these remembered features. Synthesized trajectories generated by using sequences of feature-based control primitives approximately fit the mean behavior of the bats. However, as noted in Section \ref{section:statistical}, bats following leaders seem to behave differently form those that do not. The interaction between the bats is a factor that the previous work of \cite{kong2013optical} does not consider. In this section, we discuss a strategy that takes the bat-bat interaction into consideration and show that now the statistics, both the mean and the variance, of the synthesized trajectories fit with those of the bat data.

\subsection{Is Pure Following Strategy Sufficient?}
\label{s:pure_following}

\begin{figure}[!tbh]
\centering
\includegraphics[width=\columnwidth]{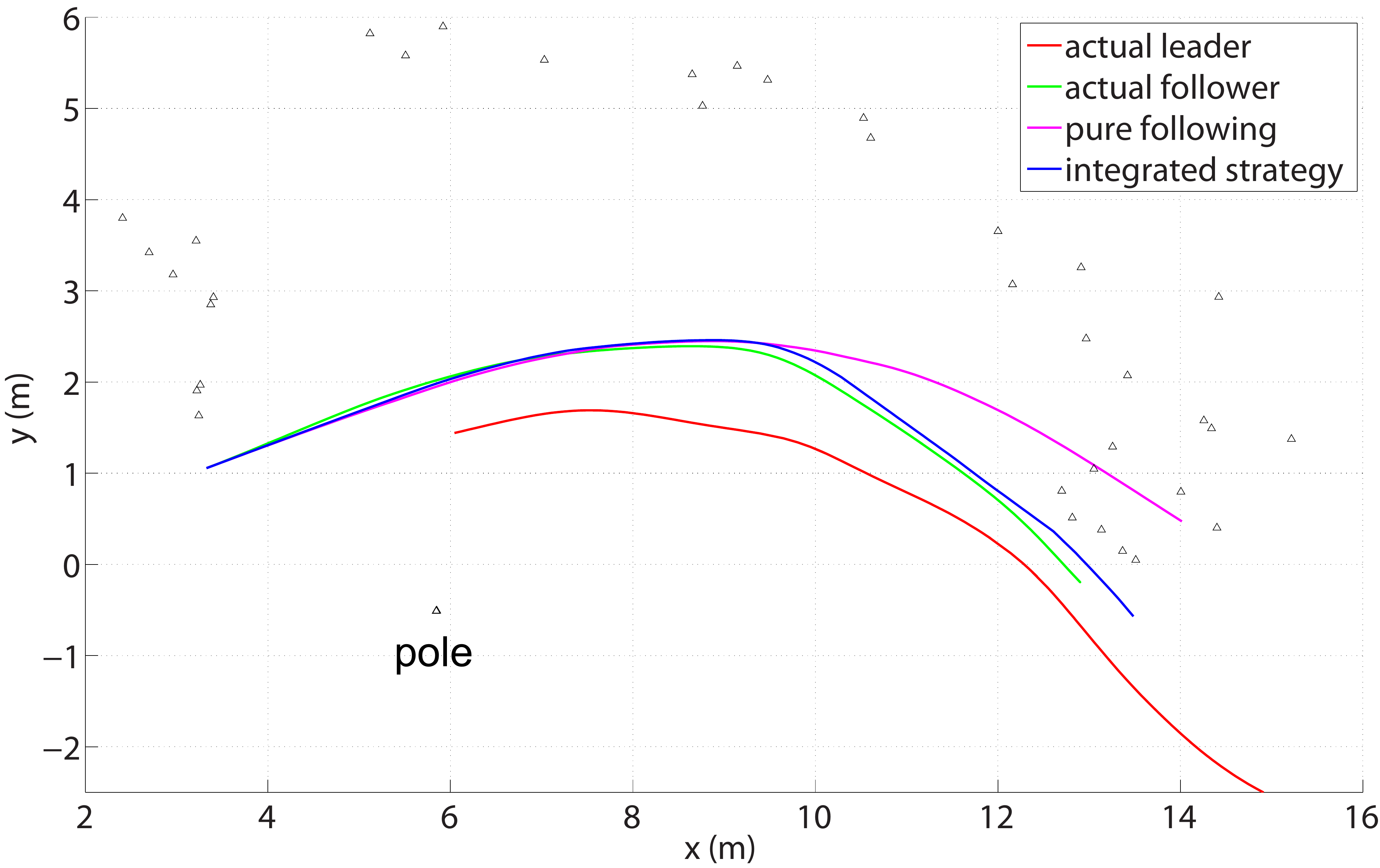}
\caption{Actual bat trajectories (red: leader bat, green: follower bat) and synthesized follower trajectories (purple: based on control law (\ref{e:thm_control_tau}), blue: based on the integrated strategy).}
\label{fig:Simulation_Data}
\end{figure}

Fig. \ref{fig:Simulation_Data} shows the actual trajectories of a leader-follower pair and a synthesized follower trajectory (purple) by using control law (\ref{e:thm_control_tau}) with the assumption that the follower only reacts to the leader without utilizing either its spatial memory or cues from the environment. The purple synthesized trajectory fits with the actual follower bat's trajectory (green) well for the segment that has $x$ coordinates smaller than 9 meters, which implies that the follower bat synchronizes its motion with the leader inside the open space between the pole and the wooded area. After passing 9 meters, the discrepancy between the synthesized and actual trajectories becomes larger. The synthesized trajectory has the danger of colliding with the obstacles or losing track of the leader due to occlusion. Here, we need to consider a navigation strategy that integrates a rapid refocus of attention on the looming tree obstacles.

\subsection{Integrated Strategy: Spatial Memory Fused with \newline Reactions to Environment and Other Bats}

\begin{figure}[!tbh]
\centering
\includegraphics[width=\columnwidth]{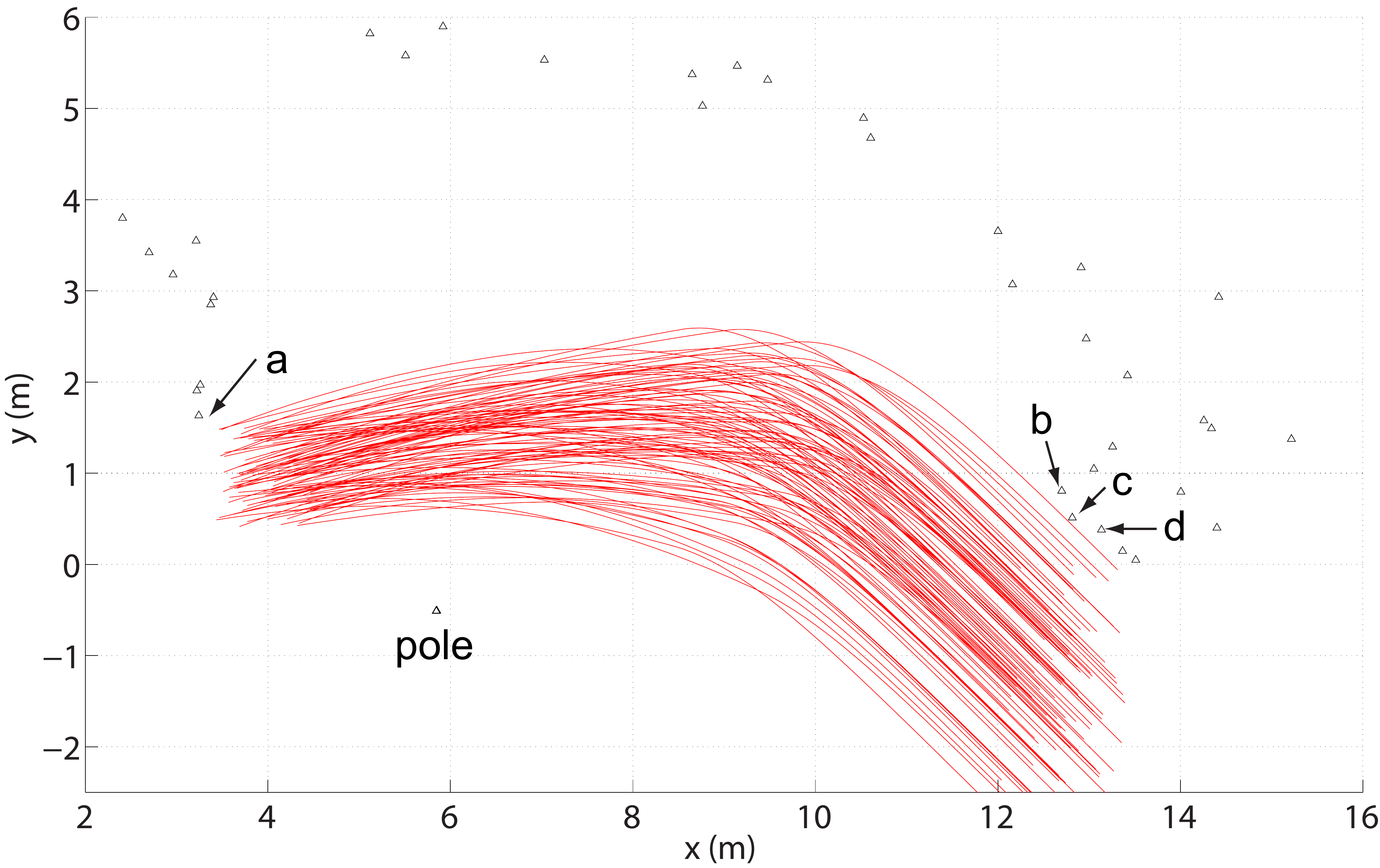}
\caption{100 synthesized trajectories based on the new integrated strategy. Labeled features are the ones that are assumed being memorized by the bats.}
\label{fig:Final_Simulation}
\end{figure}

\begin{figure}[!tbh]
\centering
\includegraphics[width=\columnwidth]{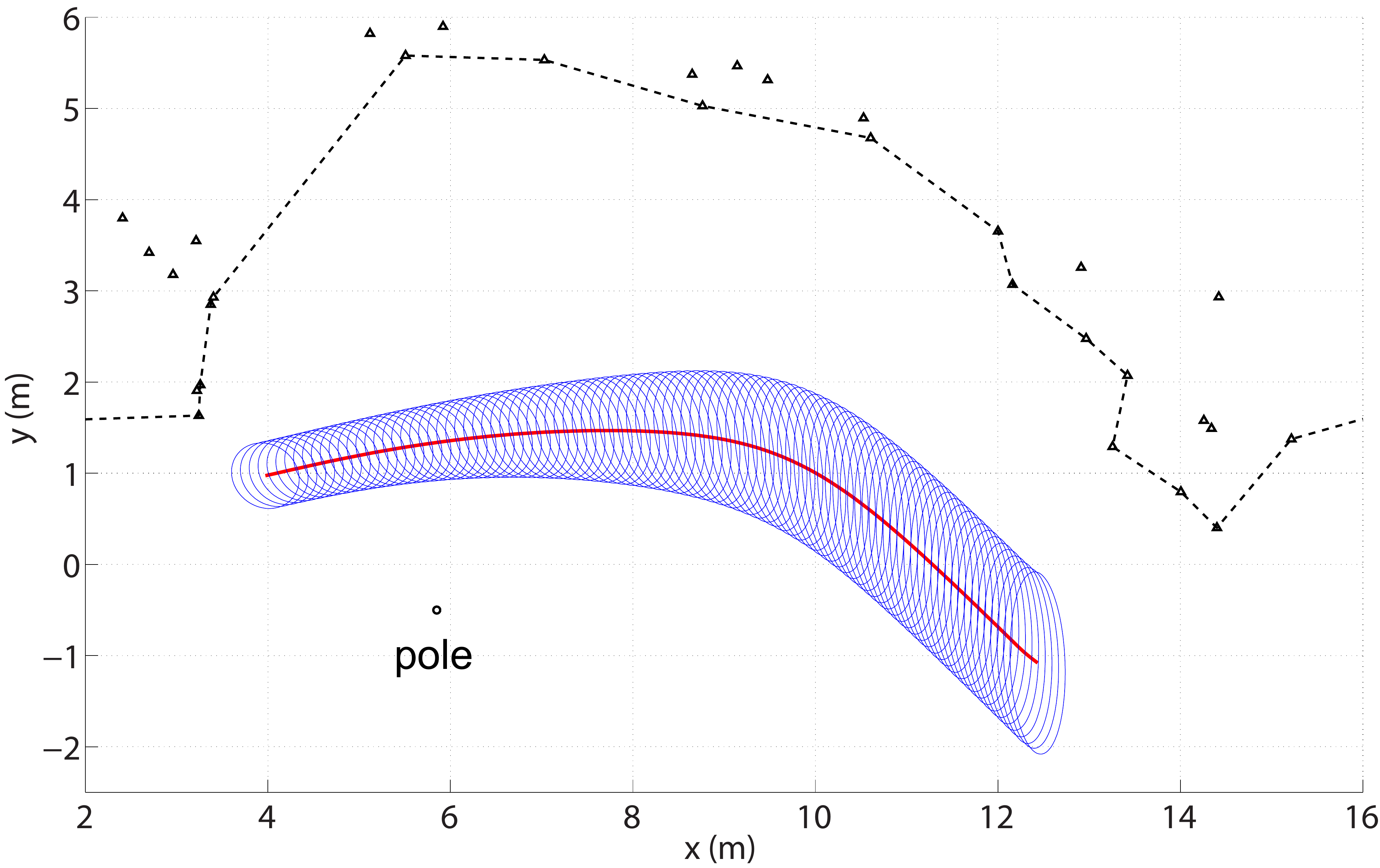}
\caption{Flight path statistics of the 100 synthesized trajectories depicted in Fig. \ref{fig:Final_Simulation}. The red curve is the mean trajectory, and the blue ellipses (centered on the mean trajectory) represent a dispersion of one standard deviation.}
\label{fig:Ellip_2}
\end{figure}

The integrated strategies proposed in \cite{kong2013optical} are now extended so as to incorporate the following behavior. Navigation strategies are synthesized from three vision-based control primitives: a distance maintenance law $u_d[\mathcal{O}_1,\mathcal{O}_2]$, a circling control law $u_c[\mathcal{O}_1]$ and a following control law $u_f[\mathcal{O}_1]$, where $\mathcal{O}_1$ and $\mathcal{O}_2$ are environmental features used in a particular control law and can be either static (for $u_d$ and $u_c$) or moving (for $u_f$). The primitives $u_d$ and $u_c$ can be found in \cite{kong2013optical}, while the primitive $u_f$ is Eq. (\ref{e:thm_control_tau}). 

Fig. \ref{fig:Final_Simulation} shows 100 synthesized trajectories based on the new integrated strategy. The vehicles are assumed to move according to Eqs. (\ref{e:leader}) and (\ref{e:follower}) with a constant speed. The vehicles appear in the field in accordance with a Poisson process. Their arrival locations and velocities are generated randomly by a Gaussian model with its mean and variance the same as those of the collected bat data. (We only simulate the bats' behavior after they pass feature `a' as shown in Fig. \ref{fig:Final_Simulation}.) The intersubjective distance between a pair of vehicles determines whether there exists a leader for the trailing vehicle to follow. If there exists a leader, the follower vehicle relies on the leader and the control $u_f$ for navigation. It switches to environment-cue-directed control $u_d$ or $u_c$ when it is on a collision course. On the other hand, if there does not exist a leader, the follower vehicle relies on its spatial memory and cues from the environment for navigation and the controls they can use are $u_d$ and $u_c$. For Fig. \ref{fig:Final_Simulation}, each trajectory is generated by a sequence of controlled motion segments as follows:
\begin{itemize}
\item If there does not exist a leader, the trajectory is generated by $u_c[\text{pole}] \rightarrow u_d[b, c] \rightarrow u_d[c, d] \rightarrow u_d[.,.]$ for the reamining features;
\item If there exits a leader, the trajectory is generated by $u_f[\text{leader}] \rightarrow u_d[b, c] \rightarrow u_d[c, d] \rightarrow u_d[.,.]$ for the reamining features. 
\end{itemize}
For Fig. \ref{fig:Simulation_Data}, the follower trajectory (blue) is generated by the second strategy since it has a leader (red). We prescribe the switching between the primitives based on the nearest feature(s) in the follower's body $x_f$ direction. For instance, the switching from $u_f[\text{leader}]$ to $u_d[b, c]$ is triggered if feature $b$ is closer to the follower than the leader in the $x_f$ direction. Similarly, the switching from $u_d[b, c]$ to $u_d[c, d]$ is triggered if feature $d$ is closer to the follower than feature $b$ in the $x_f$ direction.

The statistics for the 100 trajectories are shown in Fig. \ref{fig:Ellip_2}. A comparison between Fig. \ref{fig:Ellip_2} and Fig. \ref{fig:Variance1new} shows that the synthesized trajectories accurately capture both the mean and the variance of the actual bat trajectories with the only difference being that the ellipses in Fig. \ref{fig:Variance1new} are slightly fatter. One possible explanation of the difference is that the sensors are assumed to be noiseless for the synthesized trajectories while this is not the case for actual bats. Similarity can also be observed between the actual follower's trajectory (green) and the synthesized trajectory based on the integrated strategy (blue) in Fig. \ref{fig:Simulation_Data}. Such resemblances support our integrated strategy hypothesis.

By following another bat, in the context of navigation, a follower bat can save energy by adopting a low duty cycle echolocation or even turning off its sonar completely (\cite{chiu2008flying}). It can also be used by an inexperience individual to follow an experienced one. In such case, the leader (e.g. a female bat) is more familiar with the environment than the follower (e.g. a juvenile). It is important to note that relying solely on following is not a robust strategy. Occasionally the follower bat needs to sense the environment in order to update its spatial memory and avoid collisions. 

\subsection{Robotic Implementation}

\begin{figure}[!tbh]
\centering
\subfigure[]{\includegraphics[width=0.9\columnwidth]{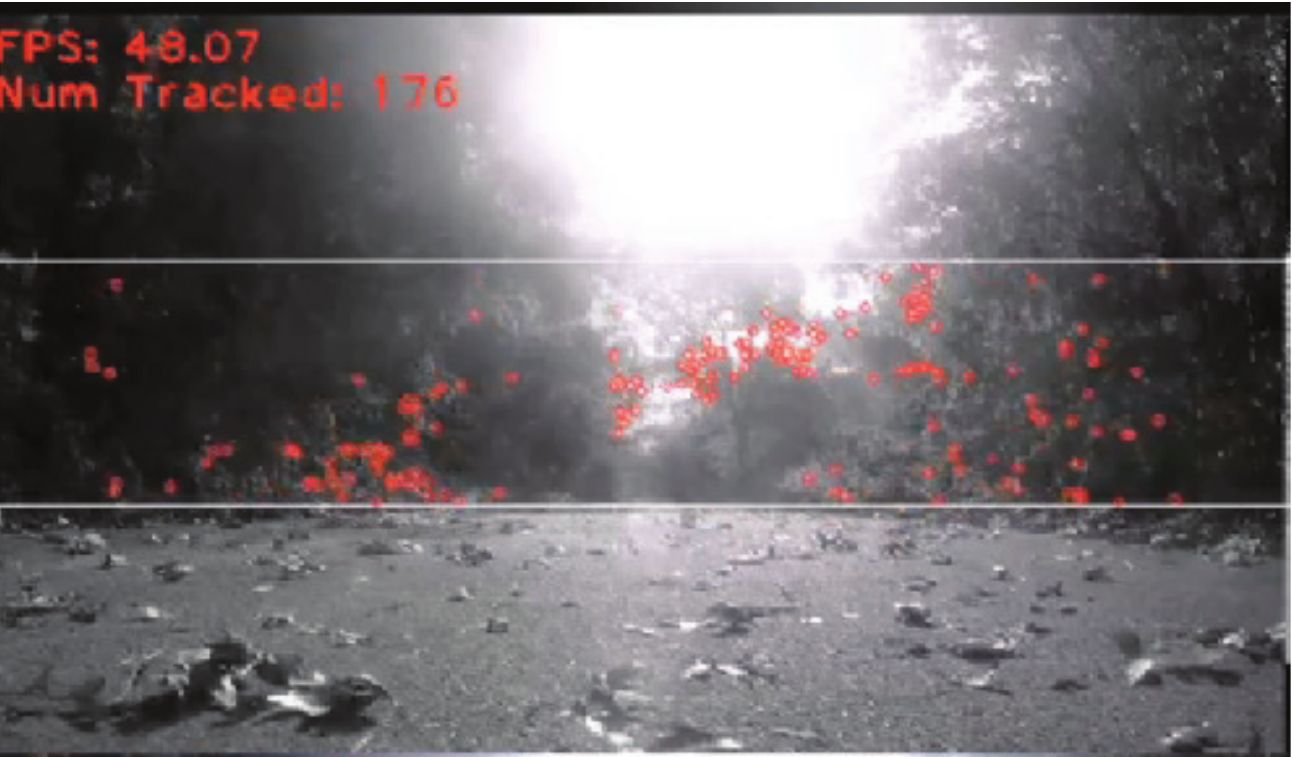}\label{sub:Loom_1}}
\subfigure[]{\includegraphics[width=0.9\columnwidth]{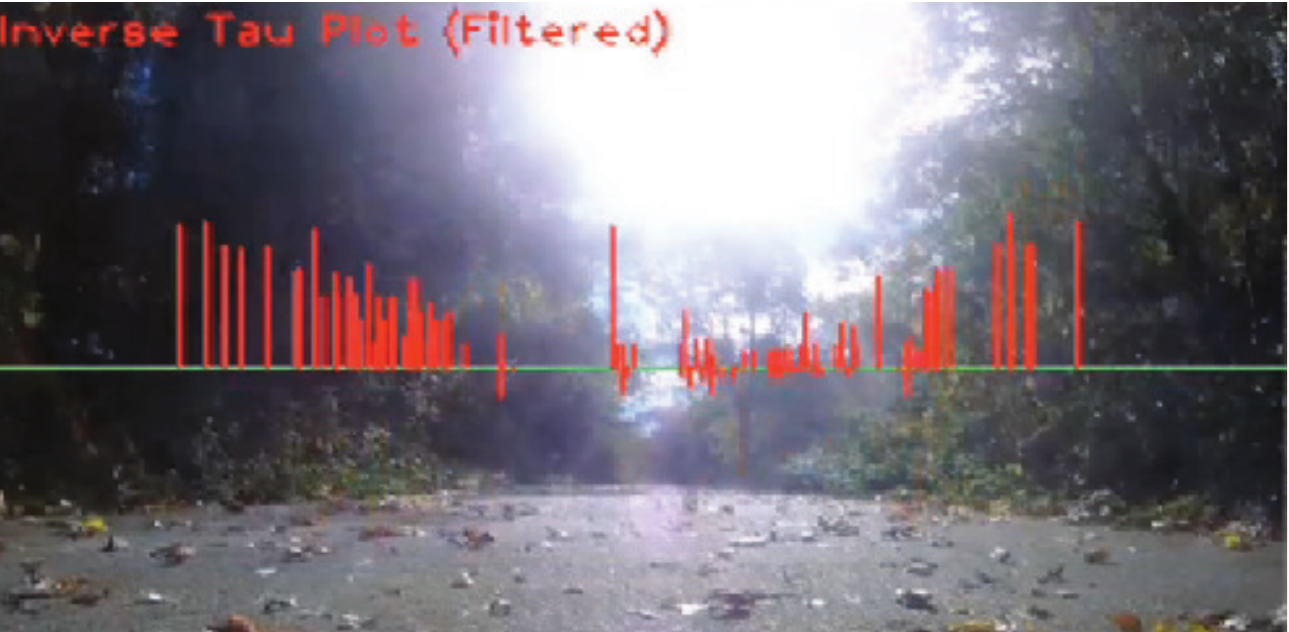}\label{sub:Loom_2}}
\subfigure[]{\includegraphics[width=0.9\columnwidth]{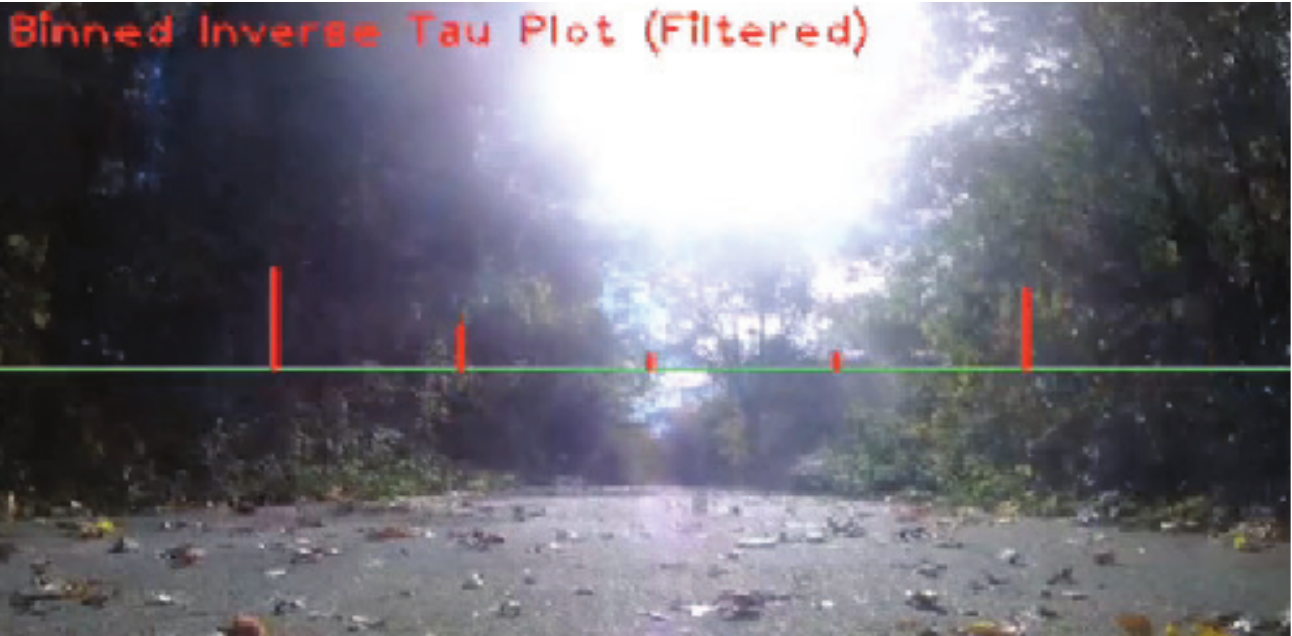}\label{sub:Loom_3}}
\caption{Optical flow implementation results. (a) Circles indicate locations of features selected by the FREAK algorithm. (b) $\lambda$ computed by using optical flow derived from the Lucas-Kanade algorithm. (c) Binned $\lambda$.} \label{fig:Loom}
\end{figure}

Our lab (Intelligent Mechatronics Lab at Boston University) is currently implementing the knowledge we have learned from the bats to the navigation and control of autonomous vehicles. Fig. \ref{fig:Loom} illustrates some of the implementation results with a ground rover mounted with a single camera. Circles in Fig. \ref{sub:Loom_1} are features selected by the FREAK (Fast Retina Keypoint) algorithm (\cite{alahi2012freak}). They correspond to the features that are utilized by the bats for their navigation. Optical flow can be computed for these feature points by using the Lucas-Kanade algorithm (\cite{lucas1981iterative}) and then be used to compute loom $\lambda$ for each feature, which is shown in Fig. \ref{sub:Loom_2}. Due to the noisy nature of the sensors, as reflected by the negative values of $\lambda$ shown in Fig. \ref{sub:Loom_2}, a voting mechanism is implemented on the extracted loom information before feeding it to the controller as shown in Fig. \ref{sub:Loom_3}. We are also looking at ways of integrating following into our system so that our robots can, for instance, follow a person in a crowed corridor. Such an integrated system can potentially be used in service and human assistant robots.

\section{Conclusion and Future Work}
\label{s:sconclusion}

In this paper, we analyze a set of \emph{M. velifer} trajectories collected from field observation and show that, for a pair of bats that emerge successively, the flight behavior of the follower is significantly affected by that of the leader, which can not be explained by existing pursuit control laws. We propose a concept called \emph{virtual loom} $\lambda$, which captures the geometrical configuration of a bat pair. We then introduce a steering law based on $\lambda$ and show that synthesized trajectories generated by following an integrated strategy, which combines spatial memory, environment-cue-based control and leader-cue-based control and stitches together a sequence of vision-based motion primitives, exhibit behaviors that are similar to the observed bat behavior.

In this paper, the switching strategy between control primitives is prescribed by experts. A more data-driven research is being planned so as to learn from the data the switching boundaries or the switching laws. As mentioned in Remark \ref{r:new_data}, our most recent work has gathered a larger data set. This new data includes audio tracks that are synchronized with the video tracks to help better understand the role of echolocation calls. We plan to analyze them in order to get a holistic understanding of the roles of different sensory modalities in navigation and bat interactions.

\section*{Acknowledgment}

The authors would like to thank Diane Theriault and Margrit Betke from Boston University for kindly providing us the experiment data. The authors would also like to thank Ryan Shelton and Ty Hedrick from University of North Carolina for their insightful discussion regarding the paired flight behavior of swallows. 

\bibliography{references}

\appendix 

\section{Smoothing and Computation of Path Statistics}
\label{appen:computation}

Smoothing of the bat trajectories collected from the field experiment is carried out using cubic spline smoothing with a smoothing factor $F=0.85$. $F$ controls the trade-off between the fidelity to the data and the roughness of the function estimate. It is chosen such that the smoothing is good enough for noise cancellation without loosing too much information. In order to investigate features such as mean trajectory and variance along the trajectories, the smoothed trajectories are parameterized by arc length. The mean paths are calculated using the step size of $0.1$ meter interval along the arc length and the mean points are connected to each other. The variances of sample points at each arc length position are illustrated by drawing the variance ellipse by first calculating the 2-by-2 covariance matrix, its eigenvalues and eigenvectors (Fig. \ref{fig:Variance1new}, Fig. \ref{fig:leaderfollowerellipse} and Fig. \ref{fig:Ellip_2}). Eigenvectors define the orientation of the principal axes of the ellipse and the eigenvalues define the length of each principal axis. For the 2-D data, along the mean path of the trajectories the evolution of the variance ellipse is plotted using the same arc length representation and step size. In order to keep statistical significance, a threshold is set to the algorithm such that it performs the simulation for at least 20 trajectories.

\end{document}